\documentstyle[12pt]{article}

\begin{document}

\begin{titlepage}
\null\vspace{-62pt}

\pagestyle{empty}
\begin{center}

\vspace{1.0truein} {\Large\bf An application of the renormalization
                group to the calculation of the vacuum decay
                 rate in flat and curved space-time}

\vspace{1in}
{\large Dimitrios Metaxas} \\
\vskip .4in
{\it Department of Physics,\\
National Technical University of Athens,\\
Zografou Campus, 15780 Athens, Greece\\
metaxas@central.ntua.gr}\\

\vspace{0.5in}

\vspace{.5in}
\centerline{\bf Abstract}

\baselineskip 18pt
\end{center}

I show that an application of renormalization group arguments may
lead to significant corrections to the vacuum decay rate for phase
transitions in flat and curved space-time. It can also give some
information regarding its dependence on the parameters of the
theory, including the cosmological constant in the case of decay in
curved space-time.

\end{titlepage}

\newpage
\pagestyle{plain}
\setcounter{page}{1}
\newpage

\section{Introduction}

The calculation of the false vacuum decay rate in a first-order
phase transition in quantum field theory is based on methods
developed in \cite{coleman1}. One proceeds by finding a bounce
solution to the classical Euclidean field equations. Then the bubble
nucleation rate per unit volume, $\Gamma$, can be written in the
form
\begin{equation}
\Gamma = A e^{-B},
 \label{standard}
\end{equation}
where $B$ is the Euclidean action of the bounce, and $A$ is an
expression involving functional determinants calculated in the
background of the bounce, which is usually considered to be of order
one times a dimensionful parameter of the theory. However, this
procedure is not valid in the cases where quantum effects become
important. For example, there are cases where even the metastability
of the false vacuum is induced by quantum or high temperature
effects. One then uses the effective potential in order to calculate
the bounce solution. The use of the running effective couplings and
 renormalization effects in $B$ leads to significant corrections
\cite{sher, linde}, as does the numerical calculation of $A$ in
various cases \cite{tetradis, baacke, baacke2} using a method
described in \cite{coleman2}. An analytical method that was
developed for the treatment of radiative corrections \cite{ejw}
leads to significant contributions to (\ref{standard}) in a
consistent manner \cite{ejw, metaxas}. The method described in
\cite{ejw} is interesting because it provides a perturbative
calculation of corrections to quantities that are non-perturbative
in nature. However, this method applies to vacuum decay at zero
temperature, in flat space-time, in theories that are not plagued by
infrared divergencies \cite{ejw, metaxas, metaxas2, tetradis}.

The situation is less clear if one wishes to include gravitational
effects \cite{CdL}. Many formal results regarding the symmetry of
the bounce and the existence of a negative eigenvalue of the
functional determinants in $A$, that exist for flat space, have only
recently  been addressed in curved space-time \cite{lavrelasvili1,
lavrelasvili2, dunne} and there are many new features \cite{HM,
balek, klee, hack} due mainly to the finiteness of the compactified
de Sitter space.

A better knowledge of the decay rate (\ref{standard}) in flat and
curved space-time is therefore important, in view of the numerous
cosmological applications (see, for example, \cite{string} and
references therein).

Here I show that the application of renormalization group arguments
to (\ref{standard}) may lead to important corrections to the
prefactor $A$, that is corrections larger than order one, and also
to some clues regarding the dependence of the bubble nucleation rate
on the parameters of the theory (masses, coupling constants). In the
case of vacuum decay with gravity these parameters include the
vacuum energy or cosmological constant, and this dependence may be
significant for cosmological applications \cite{string}.

In Sec.~2 I will apply the standard renormalization group arguments
to the calculation of the decay rate, $\Gamma$, in flat space-time.
Similar arguments have been applied for the calculation of
corrections to the effective potential and the bounce action, $B$
\cite{sher, kastening, kugo, ford, quiros, alkuwari}, they have not
however been applied to the prefactor, $A$. I show that, depending
on the parameters of the theory, these effects may become important.
In Sec.~3 I will apply these arguments in the case of vacuum decay
in curved space-time, including the renormalization group dependence
of the vacuum energy or cosmological constant \cite{kastening, kugo,
ford}. I conclude with some comments in Sec.~4.

\section{Renormalization group for vacuum decay in flat space-time}

I consider a theory with a single scalar field $\phi$ with mass $m$
and coupling $\lambda$, with an effective potential $U(\phi)$ that
has a relative minimum at $\phi=\phi_f$ and an absolute minimum at
$\phi=\phi_t$. In order for this to happen usually there are
additional fields, with additional couplings and masses, that I will
not denote explicitly. The Euclidean action
\begin{equation}
S = \int d^4x \left( \frac{1}{2}
                         (\partial\phi)^2 +U(\phi) \right)
\end{equation}
is minimized at the solution of
\begin{equation}
  \Box \phi=\frac{\partial U}{\partial\phi}
  \label{beq}
\end{equation}
with appropriate boundary conditions, which gives the bounce
solution $\phi_b (x)$. The effective potential satisfies the
renormalization group (RG) equation, for a renormalization scale
$\mu$,
\begin{equation}
{\cal D} U=4\gamma U
\end{equation}
\begin{equation}
{\cal D} =\mu\frac{\partial}{\partial\mu} +
          \beta_{\lambda}\frac{\partial}{\partial\lambda}+
          \gamma_{m} m\frac{\partial}{\partial m}
\end{equation}
with $\beta_{\lambda}, \gamma_{m}, \gamma$ the modified RG functions
\cite{ford}, with solution
\begin{equation}
U=\xi^4(t) U(\phi,\lambda(t), m(t), \mu(t))
 \label{scale1}
\end{equation}
where $\mu(t)=\mu e^t$,
\begin{equation}
\xi(t)=e^{-\int_0^t \gamma(t')dt'}
\end{equation}
and $\lambda(t), m(t)$ the running coupling and mass.

Using the scaling properties of the bounce equation (\ref{beq}) and
(\ref{scale1}) we see that the relative equation for the bounce is
\begin{equation}
\phi_b = \phi_b(\xi^2(t) x, \lambda(t), m(t), \mu(t)).
 \label{scale2}
\end{equation}
Then the bounce action becomes
\begin{equation}
S=\xi^{-4}(t) S(\lambda(t), m(t), \mu(t))
\end{equation}
and satisfies
\begin{equation}
{\cal D} S = -4\gamma S.
\end{equation}
We write a general expression for the false vacuum decay rate:
\begin{equation}
\Gamma=m^4\,A(\lambda, m, \mu)\,e^{-S(\lambda, m, \mu)}
\end{equation}
where $A$ is a dimensionless function that comes from the functional
determinants involved in the calculation of the vacuum decay rate
and $S$ is the bounce action with the false vacuum contribution
subtracted \cite{coleman1}. Then imposing ${\cal D}\Gamma =0$ gives
\begin{equation}
{\cal D} A = -4(\gamma S +\gamma_m) A.
\end{equation}
This is the equation that expresses the renormalization group
invariance of the physical quantity $\Gamma$, with solution
\begin{equation}
A=A(\lambda(t), m(t), \mu(t))e^{4\int_0^t(\gamma S +\gamma_m)dt'}.
\label{pref1}
\end{equation}
We see that the prefactor $A$ has a dependence on the parameters of
the theory that may be larger than order one, depending on $\gamma,
\gamma_m, S$. For the simple case of a $\lambda \phi^4$ theory (for
negative $\lambda$), where $\gamma_m\sim\lambda,
\gamma\sim\lambda^2, S\sim1/\lambda$, it is easy to see that the
correction is smaller than order one (turns out to be of order
$\lambda$), hence insignificant for the calculation of the decay
rate. However, for other theories where $S$ scales differently with
$\lambda$ this gives corrections that are exponentially large,
although smaller than the exponential of the bounce action. This is
the case, for example, for theories with radiative symmetry breaking
where $\gamma\sim g^2, S\sim 1/g^4$, with $g$ the gauge coupling and
the corrections turn out to be of order $1/g^2$, that is larger than
order one, although smaller than the bounce action. This is expected
from the results in \cite{ejw, metaxas}, although one does not
expect the contributions from the renormalization group to provide
the complete higher order corrections that were calculated there.

This is, however, a general argument that applies to any model, and
it can give some information concerning the higher order corrections
and their dependence on the parameters involved. In the case of
vacuum decay in curved spacetime, treated in the next Section, it is
more useful, since higher order calculations may involve quantum
gravity effects that cannot be seen in perturbation theory.

\section{Renormalization group for vacuum decay in curved
space-time}

Unlike vacuum decay in flat space-time, in the case of tunneling in
de Sitter space-time the exponent $B$ in (\ref{standard}) is the
difference of two finite actions: the bounce action and the finite
Euclidean action of de Sitter space-time that depends solely on the
cosmological constant. The different renormalization group behaviour
of these two gives additional contributions to the previous
estimates that also depend on the cosmological constant.

For definiteness I will consider a potential term (with coupling
constant $\lambda$ and mass parameter $m$) that is everywhere
positive and has two minima, a relative minimum at $\phi_f$ with
$U(\phi_f)=\Lambda$, and an absolute minimum at $\phi_t$. I will
also consider the thin wall approximation \cite{coleman1} where the
difference $\varepsilon$ between the two minima satisfies
\begin{equation}
\lambda\varepsilon << m^4.
 \label{con1}
\end{equation}
Then the flat space bounce solution has radius
\begin{equation}
R_b \sim \frac{m^3}{\lambda\varepsilon}.
 \label{rb}
\end{equation}
The curved spacetime formalism of \cite{CdL} calculates the
gravitational contributions to the decay rate for this case, which
corresponds to tunneling  from a de Sitter space with cosmological
constant $\Lambda$  to a de Sitter space with a slightly smaller
cosmological constant $\Lambda-\varepsilon$.

In order to apply the dilute instanton gas arguments of
\cite{coleman1, coleman2}, however, we need to work in the
approximation where the bounce radius is much smaller than the
radius of the compactified de Sitter space
\begin{equation}
R_{dS} = \left(\frac{3M_P^2}{8\pi\Lambda}\right)^{1/2}.
 \label{rds}
\end{equation}
For $R_b<<R_{dS}$ the gravitational corrections to the flat space
results are of higher order. It is this approximation (together with
the thin wall approximation mentioned before) that I will use here
in order to apply the arguments of the previous Section. In this
case we can use the renormalization group coefficients for flat
space as a first approximation. Higher (gravitational) corrections
to these terms will be suppressed by powers of $R_b/R_{dS}$.

 The arguments of the previous Section can be extended to the case of false vacuum
decay in curved space-time if we include the running of the vacuum
energy in the effective potential \cite{kugo, ford}. In order to
describe tunneling in curved space-time \cite{CdL} one considers an
$O(4)$-invariant metric
\begin{equation}
ds^2=d\tau^2 + \rho(\tau)^2 (d\Omega)^2.
\end{equation}
The Euclidean equations of motion are
\begin{equation}
\phi''+\frac{3\rho'}{\rho}\phi'=\frac{\partial U}{\partial\phi}
 \label{cdl1}
\end{equation}
\begin{equation}
\rho'^2=1+\frac{8\pi G}{3}\rho^2 \left(\frac{1}{2}\phi'^2-U\right)
 \label{cdl2}
\end{equation}
where the prime is $d/d\tau$, and the Euclidean action is
\begin{equation}
S=4\pi^2\int d\tau \left(\rho^3 U-\frac{3\rho}{8\pi G}\right).
\end{equation}
We write the effective potential as $U=U_0 + \Lambda$ where $U_0$ is
the part that vanishes at $\phi_f$, the relative minimum, and
$\Lambda$ is the cosmological constant term, with
$d\Lambda/dt=\beta_{\Lambda}$. As was mentioned before, I will
consider the flat space value for $\beta_{\Lambda}$ \cite{kugo,
ford}, which is of order $m^4$, and neglect the gravitational
corrections to $\beta_{\Lambda}$ which are expected to be of order
$R_b/R_{dS}$.

 The exponent $B$ for the tunneling
rate is $B=S_b-S_f$ with $S_b$ the bounce action and $S_f$ the false
vacuum action
\begin{equation}
S_f = -\frac{8\pi M_P^4}{3\Lambda}.
\end{equation}
In order to exploit the scaling properties of the bounce solution we
write the solution of the renormalization group equation for $U$ as
\begin{equation}
U=\xi^4(t) U(\phi, \lambda(t), m(t), \mu(t), \bar{\Lambda}(t))
\end{equation}
where $\bar{\Lambda}=\xi^{-4}\Lambda$. Then the solutions $\phi_b$,
$\rho_b$, of (\ref{cdl1}) and (\ref{cdl2}) scale as
\begin{equation}
 \phi_b=\phi_b(\xi^2(t)\tau, \lambda(t), m(t), \mu(t), \bar{\Lambda}(t))
\end{equation}
\begin{equation}
\rho_b=\frac{1}{\xi^2(t)}\rho_b (\xi^2(t) \tau, \lambda(t), m(t),
\mu(t), \bar{\Lambda}(t))
\end{equation}
and the bounce action similarly to the previous case
\begin{equation}
S_b=\xi(t)^{-4} S(\lambda(t), m(t), \mu(t), \bar{\Lambda}(t))
\end{equation}
\begin{equation}
{\cal D} S_b = -4\gamma S_b(\bar{\Lambda}),
\end{equation}
where ${\cal D}$ includes the running of $\bar{\Lambda}$.  For the
false vacuum action we write
\begin{equation}
\frac{d}{dt} S_f =-\frac{8\pi M_P^4}{3 \bar{\Lambda}}
                  \left(4\gamma -4\gamma -\frac{1}{\xi^4\Lambda}
                                \frac{d\Lambda}{dt}\right)
\end{equation}
to get
\begin{equation}
{\cal D} B= -4\gamma B_{CdL}(\bar{\Lambda})
           +\frac{8\pi M_P^4}{3\Lambda}
           \left(\frac{\beta_{\Lambda}}{\Lambda}
                 +4\gamma\xi^4 \right)
\end{equation}
where $B_{CdL}(\bar{\Lambda})$ is the exponent for decay in curved
space-time \cite{CdL}, more precisely the one derived in
\cite{parke} for arbitrary values of the cosmological constant, with
the modified running for $\Lambda$. Using similar arguments as in
the previous section we get for the prefactor
\begin{equation}
A=A(t)
    \exp\left(  4\int \left(\gamma B_{CdL}(\bar{\Lambda})
    +\gamma_m\right)dt'
    -\int\frac{8\pi M_P^4}{3 \Lambda}
       \left(\frac{\beta_{\Lambda}}{\Lambda} +4 \gamma
       \xi^4\right)dt'
       \right).
\end{equation}
We see again that this gives a correction to the prefactor which may
be smaller than the exponential but also, depending on the specifics
of the model, may be larger than order one, like in the flat case,
hence important for the calculation of the decay rate. Again, this
contribution has a non-trivial dependence on the parameters of the
theory, including the cosmological constant.

In summary our approximations are $R_b << R_{dS}$ together with the
thin wall approximation (\ref{con1}). They hold for a large range of
the parameters of the model, provided that the cosmological constant
term (and the mass parameters of the theory) are not of Planck scale
value. That is, we are not calculating quantum gravitational
contributions, some of which may be estimated by extending this
method with the renormalization group in curved spacetime \cite{gr4,
gr1, gr2, gr3}. It is interesting, however, that a non-trivial
dependence of the prefactor on the cosmological constant term arises
even at this scale.

\section{Comments}

We see that an application of the usual renormalization group
arguments can give some contributions to the calculation of the
false vacuum decay rate, especially in theories where the quantum or
gravitational effects are important. In the case of tunneling in
curved space-time this may give some insight to the dependence of
the tunneling rate on the parameters of the theory, such as the
cosmological constant. One would like to have a more detailed
description of tunneling between different vacua and show that the
most probable is the one with the observed values of the physical
parameters and the cosmological constant. The complete description
may involve contributions from quantum gravity effects that are not
treatable perturbatively. The information that can be obtained from
the renormalization group, as described in this work or otherwise,
is therefore important.

I should also note that the renormalization group arguments that I
used here are the usual ones of quantum field theory in flat
space-time, together with the running of the vacuum energy,
interpreted as a cosmological constant. They correspond, therefore,
to the limit of weak gravity, or equivalently, to the cases where
the radius of the bounce solution is much smaller than the radius of
the compactified de Sitter space, which is, anyway, the situation
where the arguments of \cite{coleman1, coleman2} can be
straightforwardly applied.
 Some ways to improve the
present results may involve the use of multi-scale renormalization
group arguments \cite{ford2, kugo2} and the use of renormalization
group in curved space-time \cite{gr4, gr1, gr2, gr3}.

\vspace{0.5in}

\centerline{\bf Acknowledgements} \noindent Most of this work was
done while visiting the National Technical University of Athens. I
am grateful to the people of the Physics Department for their
hospitality.

\end{document}